\documentclass{aa}

\def\kms {\hbox{${\rm km\ s}^{-1}$}}
\def \HI {H{\sc \,i}}

\def\scm  {$\hbox{{\rm cm}}^{-2}$}    
\def\arcsec {\hbox{$^{\prime\prime}$}}

\def\lapp{\ifmmode\stackrel{<}{_{\sim}}\else$\stackrel{<}{_{\sim}}$\fi}
\def\gapp{\ifmmode\stackrel{>}{_{\sim}}\else$\stackrel{>}{_{\sim}}$\fi}

\begin{document}

\title{A search for molecules in damped Lyman-alpha absorbers occulting
millimetre-loud quasars\thanks{based on results collected at Onsala Space
Observatory, Sweden and the European Southern Observatory, La Silla,
Chile}}

\author{S. J. Curran\inst{1}, M. T. Murphy\inst{1}, J. K. Webb\inst{1},
F. Rantakyr\"{o}\inst{2,3}, L. E. B. Johansson\inst{4} and
S. Nikoli\'{c}\inst{4,5}}

\offprints{S. Curran}

\institute{School of Physics, University of New South Wales, Sydney NSW
2052, Australia \and European Southern Observatory, Casilla 19001, Santiago
19, Chile \and Observatorio Cerro Calan, Universidad de Chile, Santiago,
Chile \and Onsala Space Observatory, Chalmers University of Technology, S
439 92 Onsala, Sweden \and Astronomical Observatory, Volgina 7, 11160
Belgrade, Serbia \\email: sjc@bat.phys.unsw.edu.au\\ }

\date{Received ; accepted }

\abstract{We have used the SEST 15-metre and Onsala 20-metre telescopes to
perform deep (r.m.s. $\ga30$ mJy) integrations of various molecular
rotational transitions towards damped Lyman-alpha absorption systems (DLAs)
known to occult millimetre-loud quasars. We have observed 6 new systems and
improved the existing limits for 11 transitions.  These limits may be
approaching the sensitivities required to detect new systems and we present
a small number of candidate systems which we believe warrant further
observation.
 }

\authorrunning{S. J. Curran et. al.}
\titlerunning{Molecules in damped Lyman-alpha absorbers}

\maketitle

\section{Introduction}\label{sec:intro}

Millimetre-band molecular absorption systems along the line-of-sight
toward quasars can provide a powerful probe of cold gas in the early
Universe. Wiklind \& Combes (1994; 1995; 1996b) have used such
absorption lines to study a variety of properties of the absorbers
themselves (e.g. relative column densities, kinetic and excitation
temperatures, filling factors). Constraints on the cosmic microwave
background temperature can also be obtained by comparing the optical
depths of different rotational transitions (e.g. Wiklind \& Combes
1996a). If the background quasar is gravitationally lensed, time delay
studies can yield constraints on the Hubble constant (e.g. Wiklind \&
Combes 2001). However, these studies have so far been limited by the
paucity of mm-band molecular absorbers: only 4 such systems are
currently known--towards TXS 0218+357 ({Wiklind} \& {Combes} 1995), PKS
1413+135 ({Wiklind} \& {Combes} 1997), TXS 1504+377 ({Wiklind} \&
{Combes} 1996c) and PKS 1830--211 ({Wiklind} \& {Combes} 1998).

Recent attention has focused on using molecular absorption lines as a probe
for possible changes in the fundamental constants on cosmological
time-scales. Detailed studies of the relative positions of heavy element
optical transitions compared with laboratory spectra favour a smaller fine
structure constant ($\alpha \equiv e^2/\hbar c$) at redshifts $0.5 < z <
3.5$ at the 4.1$\sigma$ significance level (Murphy et al. 2001b, Webb et
al. 2001). The observed fractional change in $\alpha$ [$\Delta\alpha/\alpha
= (-0.72 \pm 0.18)\times 10^{-5}$] is very small, so systematic errors have
to be carefully considered. However, a thorough search for systematics has
not revealed a simpler explanation of the optical results (Murphy et
al. 2001c) and so independent constraints at similar redshifts are
required.

Comparison of molecular rotational (i.e. mm-band) and corresponding
H{\sc \,i}\,21-cm absorption line frequencies has the potential to
constrain changes in $\alpha$ with a fractional precision $\sim
10^{-6}$ per absorption system. The ratio of the hyperfine (\HI)
transition frequency to that of a molecular rotational line is
$\propto y \equiv \alpha^2g_p$ and so any variation in this will be
observed as a difference in the apparent redshifts of these lines
(Drinkwater et al. 1998).
 
From spectra of two of the known high redshift absorbers, PKS\,1413+135
and TXS\,0218+357, Carilli et al. (2000) and Murphy et al. (2001a) have
\begin{table*}
\begin{center}
\caption[]{{\bf Top}: The northern sample. The radio flux densities at
various GHz frequencies of the background quasars are given (see
Curran {\rm et~al.} 2002 for details). Note that a ``$\approx$''
denotes a variable flux density. $z_{\rm abs}$ is the DLA redshift and
the final column gives the molecules most commonly detected in the
four known absorbers (see Wiklind \& Combes references), which fall
into the Onsala 84--116 GHz band.  ($^*$HCO$^+$ $0\rightarrow1$ in B
08279+5255 and $^*$CS $2\rightarrow3$ in B 1017+1055 are 7 mm
observations). Note that CS has so far only been detected in PKS
1830--211 ({Wiklind} \& {Combes} 1996c) but since none of the commonly
detected molecules fell into the band at the DLA redshift we tuned to
this dense gas tracer. {\bf Bottom}: The southern sample. The final
column lists the molecules most commonly detected which fall into the
SEST 78--116 GHz and 128--170 GHz bands. }
\label{t1}
\begin{tabular}{l c c c  c r r r c c c c}
\hline\noalign{\smallskip}
Quasar & \multicolumn{2}{c}{Coordinates (J2000)} & \multicolumn{6}{c}{Radio flux densities (Jy)} &$z_{\rm abs}$ & Transition \\
 & h ~m ~s & d~~~$'$~~~$''$	&  $S_{1.4}$&  $S_{5.0}$ & $S_{22}$ & $S_{90}$&  $S_{230}$ &  $S_{350}$ &  & \\
\noalign{\smallskip}
\hline\noalign{\smallskip}
B 0235+1624 & 02 38 38.9 & 16 36 59 & 2.36& 1.64 & 2.48& $\approx1-6$& $\approx1-4$& --& 0.523869&  CS $2\rightarrow3$\\
B 0248+430 & 02 51 34.5& 43 15 16& 1.43& 0.66&  $\approx0.7$& 0.32& 0.08& --& 0.3939& CS $2\rightarrow3$\\
B 0738+313&  07 41 10.7& 31 12 00 & 2.05& --& $\approx1.3$& 0.27& 0.10&--& 0.2212&  CO $0\rightarrow1$\\
B 0827+243 &  08 30 52.1 & 24 11 00 & 0.84 &0.89& $\approx1.3$& 2.4&  $\approx1.3$&   --& 0.5247& CS $2\rightarrow3$ \\
B 08279+5255 & 08 31 41.6 & 52 45 18 &$0.001$  & --& --& --& --& 0.08& 2.97364& $^*$HCO$^+$ $0\rightarrow1$ \\
... &... & ...& ...& ...& ...&... & ...&... &... & CO $2\rightarrow3$  \\
B 1017+1055 & 10 20 08.8 & 10 40 03 & 0.22 & -- &--& --& 0.004& --& 2.380& $^*$CS $2\rightarrow3$\\
... &... & ...& ...& ...& ...&... & ...&... &... &CO $2\rightarrow3$\\
B 1328+307 & 13 31 08.3 & 30 30 33& 14.7& 1.41& 2.7& 0.75& 0.33& --& 0.69218&  HCO$^+$ $1\rightarrow2$ \\
... &... & ...& ...& ...& ...&... & ...&... &... &  CS $2\rightarrow3$\\
\noalign{\smallskip}
\hline
\noalign{\smallskip}
B 0458--020 & 05 01 12.8 & -01 59 14 & 2.2 & $\approx3$ & -- & 0.8 & $\approx0.5$ & -- &2.0399&  HCO$^+$  $2\rightarrow3$ \\
... &... & ...& ...& ...& ...&... & ...&... &... &CO $3\rightarrow4$  \\
B 0834--201 & 08 36 39.2 & -20 16 59 & 1.97 & 1.5 & --& 0.85 & 0.28 & -- & 1.715 & HCO$^+$  $2\rightarrow3$ \\
... &... & ...& ...& ...& ...&... & ...&... &... &CO $3\rightarrow4$  \\
B 1229--0207 & 12 32 00.0 & -02 24 05 & 1.90 & 0.90 & $\approx1$ & $\approx0.4$ & $\approx0.2$ & -- & 0.3950 & CO $0\rightarrow1$ \\
... &... & ...& ...& ...& ...&... &... &... &... &CO $1\rightarrow2$ \\
B 1451--375 & 14 54 27.4 & -37 47 33 & 1.57 & 1.84 & -- & $\approx1.5$ & $\approx0.5$ & -- & 0.2761 & CO $0\rightarrow1$ \\
... &... & ...& ...& ...& ...&... &... &... &... &HCO$^+$ $1\rightarrow2$ \\
\noalign{\smallskip}
\hline\end{tabular}
\end{center} 
\end{table*}
\begin{table*}
\begin{center}
\caption[]{Molecular lines previously searched for in
radio-illuminated DLAs. All of the observations were performed with
the IRAM 30-m telescope, except for HCO$^+$ $2\rightarrow3$ towards B
0458--020 which was observed with the SEST. Again, apart from those
marked ``*'' which are from {Wiklind} \& {Combes} (1995), the radio
flux densities are from the various sources cited in Curran {\rm
et~al.} (2002). $\sigma$ is the lowest r.m.s. noise obtained for this
frequency (with the reference given) recalculated for a channel width
of 10 \kms (see footnote 3). Note that {Wiklind} \& {Combes} (1994b) is a search for CO
emission, although any absorption features at the appropriate redshift
should be apparent.}
\label{wc}
\begin{tabular}{l cl cccc c c l}
\hline\noalign{\smallskip}
Quasar & $z_{\rm abs}$ & Transition & \multicolumn{5}{c}{Radio flux densities (Jy)}&  $\sigma$ [mK] & Reference \\
& & &  $S_{1.4}$&  $S_{5.0}$ & $S_{22}$ & $S_{90}$&  $S_{230}$ & & \\
\noalign{\smallskip}
\hline\noalign{\smallskip}
B 0235+1624 &0.5238  &  CO $1\rightarrow2$ & 2.36 & 1.64& 2.48& 1.2$^*$& 1.3$^*$& 2.3& {Wiklind} \& {Combes} (1995)\\
        &       & HCO$^+$ $3\rightarrow4$ & ... &... & ...& ...& ...& 3.6&  {Wiklind} \& {Combes} (1995)\\
B 0458--020 & 2.0397 & CO $2\rightarrow3$ & 2.2 & $\approx3$& --& 0.8& $\approx0.5$& 3.0&{Wiklind} \& {Combes} (1994b) \\
	& & HCO$^+$ $2\rightarrow3$ & ... & ...& ...& ...&... & 7.1&{Wiklind} \& {Combes} (1996b) \\
B 0528--2505 & 2.1408 & CO $2\rightarrow3$ &1.50 &1.13   & --& --& --& 1.3&{Wiklind} \& {Combes} (1994b)\\
B 0834--201 & 1.715 & HCO$^+$ $2\rightarrow3$ & --& 1.5& --& 0.85& 0.28& 3.6 &{Wiklind} \& {Combes} (1996b)\\
	& & HCO$^+$ $3\rightarrow4$ & ... &... &... &... &... & 3.9&  {Wiklind} \& {Combes} (1996b)\\
B 1215+333 & 1.9984 & CO $2\rightarrow3$ &0.18 & 0.08 & --& --& --& 4.4 &{Wiklind} \& {Combes} (1994b) \\ 
B 1229--0207 & 0.39498 & CO $1\rightarrow2$ & 1.90 & 0.90 & $\approx1$ & $\approx0.4$ & $\approx0.2$ & 12 & {Wiklind} \& {Combes} (1995)\\
B 1328+307 & 0.69215 &  CO $1\rightarrow2$ & 14.7 & 1.41& 2.7& 0.6$^*$&0.24$^*$ & 6.1&{Wiklind} \& {Combes} (1995) \\
	& & CO $2\rightarrow3$ & ... & ...&... &... &... & 6.5 &{Wiklind} \& {Combes} (1995) \\
	& & HCO$^+$ $1\rightarrow2$ &...  &... &... &... &... & 5.8&{Wiklind} \& {Combes} (1995) \\
B 2136+142 & 2.1346 &  CO $2\rightarrow3$ & -- & 1.11& 1.6& 0.59$^*$& 0.25& 2.8& {Wiklind} \& {Combes} (1996b) \\
	& & CO $3\rightarrow4$ & ... & ...& ...& ...& ...& 2.3& {Wiklind} \& {Combes} (1996b)\\
	& & CO $5\rightarrow6$ & ...  &...&... &... &... & 3.9&   {Wiklind} \& {Combes} (1996b)\\
	& & HCO$^+$ $2\rightarrow3$ & ... & ...&... &... & ...& 3.4& {Wiklind} \& {Combes} (1995) \\
\noalign{\smallskip}
\hline
\end{tabular}
\end{center}
\end{table*}
obtained constraints on $\Delta y/y$ consistent with zero
$y$-variation at redshifts $z_{\rm abs} = 0.6847$ and 0.24671,
respectively. However, the major uncertainty in the mm/H{\sc \,i}
comparison is that intrinsic velocity differences between the mm and
H{\sc \,i} absorption lines can be introduced if the lines-of-sight to
the millimetre wave and radio continuum emission regions of the quasar
differ, as is certainly the case for PKS\,1413+135 and TXS\,0218+357
(Carilli et al. 2000). Thus, a {\it statistical} sample of mm/H{\sc
\,i} comparisons is required to independently check the optical results.

One systematic approach to finding new high redshift molecular absorbers is to target
high column density systems with known redshifts. A convenient sample
is therefore the damped Lyman-alpha absorbers defined to have neutral hydrogen
column densities $N_{\rm HI}\ga10^{20}$ cm$^{-2}$. We therefore
compiled a catalogue of all known DLAs and shortlisted those which are
illuminated by radio-loud quasars (Curran {\rm et~al.}
2002)\footnote{Available from http://www.phys.unsw.edu.au/$\sim$sjc/dla} Of these, seven in the
northern sky have measured millimetre fluxes, and in
the south there are around a dozen DLAs illuminated by millimetre-loud
quasars, of which we observed the four loudest (Table \ref{t1}).

We note that Wiklind \& Combes (1994b; 1995; 1996b)
searched with null results for redshifted molecular emission and absorption towards 12
and 46 quasars, respectively. However, only 11 are occulted by DLAs and
not all of these are radio-loud (Table \ref{wc}). This
motivates a more systematic search for millimetre absorption 
systems associated with DLAs. In
this paper we present the results of our first search: the DLAs which
occult known millimetre-loud quasars with the SEST\footnote{The
Swedish-ESO Sub-millimetre Telescope is operated jointly by ESO and
the Swedish National Facility for Radio Astronomy, Onsala Space
Observatory, Chalmers University of Technology.} and Onsala 20-m
telescopes.

\section{Observations}

The northern 3 mm and 7 mm observations were performed in February and
April 2002, respectively, with the 20-m telescope at Onsala Space
Observatory, Sweden. For the 3 mm observations the SIS receiver was tuned to
single-sideband mode and the backend was a hybrid correlator with a
bandwidth of 1280 MHz and a channel separation of 0.8 MHz. We used
dual-beam switching with a throw of $12\arcmin$ in azimuth, and
pointing errors were typically $3\arcsec$ r.m.s. on each axis. The
intensity was calibrated using the chopper-wheel method and typical
system temperatures, on the $T_A^*$-scale, were around 300 K.  For the
7 mm observations we used the 43 GHz SIS receiver and again the backend
was the hybrid correlator with a bandwidth of 640 MHz and a
channel separation of 0.4 MHz.  Since there is no beam switch
capability at these frequencies, we removed the ripple caused by
standing waves by subtracting the Fourier components in a transform of
the spectra. Typical system temperatures, on the $T_A^*$-scale, were
around 200 K.

The southern sources in the sample were observed in April 2002 with
the 15-m SEST at La Silla, Chile, using the 100 GHz and 150 GHz SESIS
receivers. These were tuned to single-sideband mode and typical system
temperatures, on the $T_A^*$-scale, were $\approx200$~K for the RX100
and $\approx250$ K for the RX150. The backends were acousto-optic
spectrometers with 1440 channels and a channel width of 0.7~MHz. As
with the Onsala 3 mm observations, we used dual-beam switching with a
throw of about $12\arcmin$ in azimuth, and pointing errors were
typically $2\arcsec$ r.m.s. on each axis. Again, the intensity was calibrated
using the chopper-wheel method.

\section{Results and Discussion}

Upon the removal of a low order baseline and subsequent averaging of
the data for each quasar, no absorption features of $\geq3\sigma$/channel
were found\footnote{See http://www.phys.unsw.edu.au/$\sim$sjc/dla-fig1.ps.gz
for the spectra and corresponding r.m.s. noise levels.}. In Table \ref{sum}
we summarise the derived upper limits together with the previously published results.
\begin{table*}
\caption[]{Summary of our (Table \ref{t1}) and the
previously published results (Table \ref{wc}). $V$ is the visual
magnitude of the background quasar, $N_{\rm HI}$ [\scm] is the DLA
column density from the Lyman-alpha line and $\tau_{\rm 21~cm}$ is the
optical depth of the redshifted 21 cm \HI ~line (see Curran {\rm
et~al.} 2002). The optical depth of the relevant millimetre line is
calculated from $\tau=-\ln(1-3\sigma_{{\rm rms}}/S_{{\rm cont}})$,
where $\sigma_{{\rm rms}}$ is the r.m.s. noise level at a given
resolution and $S_{{\rm cont}}$ is the continuum flux density,
estimated from the values at the neighbouring frequencies (Tables
\ref{t1} and \ref{wc}). This is done for
resolutions of 10 \kms~ [$\tau_{\rm mm}(10)$] and 1 \kms ~($\tau_{\rm
mm}$), where for the latter we quote only the best existing limit. For
all optical depths, $3\sigma$ upper limits are quoted and ``--''
designates where $3\sigma>S_{{\rm cont}}$, thus not giving a
meaningful value for this limit. Blanks in the $\tau_{\rm 21~cm}$
field signify that there are no published \HI ~absorption data for
these DLAs. The final column gives the best existing limit of the
column density per unit velocity estimated for the transition [\scm
(\kms)$^{-1}$] (see main text).  }
\label{sum}
\begin{center}
\begin{tabular}{l c c c c c c c c }
\hline\noalign{\smallskip}
DLA & $z_{\rm abs}$ & Transition &  V & $N_{\rm HI}$ & $\tau_{\rm 21~cm}$ & $\tau_{\rm mm} (10)$ & $\tau_{\rm mm}$ & $N_{\rm mm}/\Delta v$\\
\noalign{\smallskip}
\hline\noalign{\smallskip}
B 0235+1624 & 0.52398$^*$ & CO $1\rightarrow2$ & 15.5 & $4\times10^{21}$ & 0.05--0.5$^a$ &  $<0.03$ & $<0.09$& $<3\times10^{14}$\\
...	&  ...& HCO$^+$ $3\rightarrow4$ & ... &... &...  & $<0.03$ & $<0.3$ & $<4\times10^{12}$\\
...& 0.523869& CS $2\rightarrow3$ & ... & ...&  ...&  $<0.2$& $<0.9$ & $<2\times10^{13}$\\
B 0248+430 & 0.3939& CS $2\rightarrow3$ & 17.7 &$4\times10^{21}$ &  0.20& $<3$ & --&-- \\
B 0458--020 & 2.0397$^*$/9  & HCO$^+$  $2\rightarrow3$  & 18.4 & $5\times10^{21}$ &0.30$^b$  & $<0.3$/$<0.2$ & $<0.4$&  $<2\times10^{12}$\\\
...	&  2.0397$^*$ & CO $2\rightarrow3$ & ... & ...&  ... & $<0.4$ & --& --\\
...	&   2.0399 & CO $3\rightarrow4$& ... & ...& ... & $<0.1$ & $<1$& $<2\times10^{16}$\\
B 0528--2505 & 2.1408 & CO $2\rightarrow3$ &19.0  &$4\times10^{20}$ & $<0.2$ & \multicolumn{3}{c}{{\it No published millimetre fluxes}} \\
B 0738+313 &  0.2212 & CO $0\rightarrow1$ & 16.1 &$2\times10^{21}$ & 0.07& $<1$ &--& --  \\
B 0827+243 & 0.5247 & CS $2\rightarrow3$ & 17.3  &$2\times10^{20}$ &0.007  &$<0.1$  & $<0.4$& $<7\times10^{12}$\\
B 08279+5255  & 2.97364& HCO$^+$ $0\rightarrow1$ & 15.2  &$1\times10^{20}$ &  &  --$^c$ &--& --\\
...	&  ... & CO $2\rightarrow3$ & ... &... & ... & --$^c$ &--& --\\
B 0834--201 & 1.715 & HCO$^+$  $2\rightarrow3$ &  18.5& $3\times10^{20}$ & & $<0.1$/$<0.2$ & $<0.4$& $<2\times10^{12}$ \\
...	&  ... &HCO$^+$  $3\rightarrow4$& ... & ...&  ...&   $<0.2$ &$<0.6$ &  $<8\times10^{12}$\\
...	&  ... &  CO $3\rightarrow4$& ... & ...&  ...&  --&--& -- \\
B 1017+1055 & 2.380& CS $2\rightarrow3$ & 17.2 & $8\times10^{19}$ &  &--$^d$ &--& --  \\
...	&  ... & CO $2\rightarrow3$ & ... & ..&  ...&  --$^d$&--& -- \\
B 1215+333 & 1.9984 & CO $2\rightarrow3$ & 18.1 & $1\times10^{21}$ &   & --$^{I,e}$ &--& --  \\
B 1229--0207 & 0.3950 & CO $0\rightarrow1$  & 16.8   &$1\times10^{21}$ &  &  $<0.3$& $<1$& $<6\times10^{15}$\\
...	&  ... & CO $1\rightarrow2$& ... & ...&  ...& $<1$  &--& --\\
...	& 0.39498$^*$  & CO $1\rightarrow2$& ... & ...&  ...&-- & --& --\\
B 1328+307 & 0.69215 & HCO$^+$  $1\rightarrow2$ & 17.3  &$2\times10^{21}$ &0.11  & $<0.2$/$<0.4$ &$<0.7$ & $<2\times10^{12}$\\
...	&  ... & CS $2\rightarrow3$ & ... &... & .... & $<0.3$ & $<2$& $<4\times10^{13}$\\
...	&  ... & CO $1\rightarrow2$ & ... & ...& ... & $<0.3$ &$<1$ &$<3\times10^{15}$\\
...	&  ... & CO $2\rightarrow3$ & ... & ...& ... & $<0.6$ & --&--\\
B 1451--375 &0.2761 & HCO$^+$ $1\rightarrow2$ & 16.7 &$1\times10^{20}$ & $<0.006$ &$<0.2$&$<0.6$ & $<2\times10^{12}$ \\
...	&  ... &CO $0\rightarrow1$  & ... & ...& ... & $<0.09$ &$<0.3$ &$<2\times10^{15}$\\
B 2136+141 & 2.1346 & HCO$^+$ $2\rightarrow3$&18.9  & $6\times10^{19}$&   &  $<0.08$ &$<0.3$ &$<1\times10^{12}$\\
...	&  ... &  CO $2\rightarrow3$ & ... &  ...&   ...&  $<0.2$ & $<0.6$& $<4\times10^{15}$\\
...	&  ... &  CO $3\rightarrow4$ & ... &  ...&   ...& $<0.2$ &$<0.8$ & $<2\times10^{16}$\\
...	&  ...&  CO $5\rightarrow6$ & ... &  ...&  ... &  $<0.6$ & --& --\\
\noalign{\smallskip}
\hline
\end{tabular}
\end{center}
{Notes: Where we have observed towards the same quasar $^*$denotes Wiklind \& Combes
results (see Table \ref{wc} for details). $^a$\HI ~absorption at $z=0.52385$
(Briggs \& Wolfe 1983). $^b$\HI ~absorption at $z=2.03945$ (Wolfe et~al. 1985;
Briggs et~al. 1989). $^c$Assumed flat spectrum, i.e.
$S_{45}=S_{87}=S_{350}=0.08$ Jy. $^d$Assumed $S_{44}\approx0.015$ Jy and $S_{102}\approx0.008$ Jy; $^e$$S_{115}\approx0.011$ Jy, based on the two available measured values (Table
\ref{t1}).}
\end{table*}
Comparing the optical depth limits with those in the four known
absorbers, the
strongest absorber, towards PKS 1830--211, has $\tau_{\rm mm}\approx1$
for HCO$^+$ $1\rightarrow2$ and $\tau_{\rm mm}\approx0.5$ for HCO$^+$
$2\rightarrow3$. The weakest, towards PKS 1413+135, has $\tau_{\rm
mm}\approx0.1$ for both CO $0\rightarrow1$ and HCO$^+$ $2\rightarrow3$. The two remaining systems have optical depths similar
to that towards PKS 1830--211 and so our new observations and the previous surveys
should have been sensitive to (at least) CO and HCO$^+$ at similar
strengths to those in 3 of the 4 known systems. Only for the
observations of B 0235+1624, HCO$^+$ $2\rightarrow3$ towards B
2136+141 (IRAM) and CO $0\rightarrow1$ towards B 1451--375 are the
surveys sensitive enough, particularly for resolutions of $\ga10$
\kms, to detect an absorption system of a strength similar to that
towards PKS 1413+135 (See Table 4).

From the optical depths we may estimate column density limits for each
transition.  In Table \ref{sum} we give the $3\sigma$ optical depth
limits according to a velocity resolution of 10 \kms ~as well as at a
resolution of 1 \kms. We give the former in order to show the limits
corresponding to a visual inspection of the spectra$^3$ and
facilitate a more direct comparison with the previous surveys.
By multiplying the latter with the expected
width of a line, the velocity integrated optical depth for a $3\sigma$
detection at $\Delta v =1$ \kms ~is obtained.  For all of the limits,
assuming LTE conditions, we can estimate the total column density of
each transition from
\begin{equation}
 N_{\rm mm}=\frac{8\pi}{c^3}\frac{\nu^{3}.f}{g_{\mu}A_{\mu}}\left.\int\right.\tau dv,
\end{equation}
where $\nu$ is the rest frequency of the $J\rightarrow J+1$
transition, $g_{\mu}$ and $A_{\mu}$ are the statistical weight and the
Einstein A-coefficient of the transition, $\int\tau dv$
is the velocity integrated optical depth and $f$ is the product of the
partition function for an excitation temperature, $T_x$ (assumed $\approx10$ K), and
$e^{E_J/kT_x}/(1-e^{-h\nu/kT_x})$ (see {Wiklind} \& {Combes} 1995, 1996b, 1999 for details).
From the optical depth limits, we derive
limits on the total column density per unit velocity, thus normalising
the limits which depend on the spectral resolution to which the data have been
smoothed.

Since we have only upper limits, and thus no knowledge of the width of
any line which may be hidden in the noise, as well as uncertainties in
the conversion to H$_2$ column densities for molecules other than CO,
it is difficult to draw meaningful comparisons between $N_{\rm HI}$
and $N_{\rm mm}$. However, by assuming an absorption line of FWHM  $\sim20$ \kms ~(as in
the case of 3 of the 4 known absorbers), we estimate for the lowest
optical depth limit (CO $1\rightarrow2$ in B 0235+1624) a value of
$N_{\rm CO}\lapp6\times10^{15}$ \scm ~at the $3\sigma$ level and a
resolution of 1 \kms, i.e. $N_{\rm CO}/N_{\rm
HI}\lapp2\times10^{-6}$. According to $N_{{\rm H}_{2}}\sim10^{4}N_{\rm
CO}$ (e.g. {Wiklind} \& {Combes} 1998), the molecular to atomic
hydrogen column density ratio is $N_{{\rm H}_{2}}\lapp2\% ~N_{\rm HI}$,
which is similar to the lower limit estimated from the optical
detection of molecular hydrogen\footnote{In the case of $z>1.8$
sources, the ultra-violet lines of H$_2$ are redshifted into the
optical window, making molecular hydrogen readily observable at these
frequencies. As well as towards B 0528--2505 ({Foltz}, {Chaffee} \&
{Black} 1988) molecular hydrogen has also been detected in the DLAs
occulting the radio-quiet quasars B 0000--2620 ({Levshakov} {\rm
et~al.} 2000), B 0013--0029 ({Ge} \& {Bechtold} 1997; {Petitjean},
{Srianand} \& {Ledoux} 2002), B 0347--3819 (Levshakov {\rm et~al.}
2002), B 1232+0815 ({Ge} \& {Bechtold} 1997; {Srianand}, {Petitjean}
\& {Ledoux} 2000) and the inferred ({Wolfe} {\rm et~al.} 1995) DLA
towards B 0551--3637 (Ledoux, Srianand \& Petitjean 2002).} in the DLA
towards B 0528--2505 ({Carilli} et al. 1996). That is, although we
have no clear detections of any molecular absorption lines, the
current limits may be approaching those sufficiently low in order
to detect molecular absorption in damped Lyman-alpha systems.

Bearing this in mind, we then analysed the spectra for tentative features
within a range corresponding to the uncertainty in the DLA
redshift\footnote{Although we have searched for possible absorption lines
close to the expected redshifts, it may possible that the line-of-sight to
the mm-band and optical emission regions of the quasars may differ
($\delta_{\rm LOS} >0$), giving rise to intrinsic offsets between any
millimetre absorption and the known DLA redshifts. Drinkwater et al. (1998)
compared millimetre and H{\sc\,i}\,21-cm Galactic absorption profiles and
found that the individual velocity components in the profiles corresponded
to within $\delta_{\rm LOS} = 1.2$ \kms. Carilli et al. (2000) argued that
$\delta_{\rm LOS} \sim 10$ \kms ~is typical of the velocity dispersion of
the interstellar medium in galaxies, and that it may even be as high as
$\delta_{\rm LOS} \sim 100$ \kms. In order to account for this, we searched
for possible absorption lines in a velocity region $[-\delta_{\rm
LOS}-\sigma_{\rm DLA}, \delta_{\rm LOS}+\sigma_{\rm DLA}]$ centered on the
expected DLA frequency with $\delta_{\rm LOS}=100$ \kms~and $\sigma_{\rm
DLA}$ the uncertainty in the DLA redshift.}. In each case the continuum
level was defined using a polynomial baseline fit and for each channel we
generated a 1$\sigma$ error from the r.m.s noise in a window of $2N_{\rm
win}+1$ channels centered on that channel. Using this error array, we
identified absorption features as series of $n$ channels over which a
deviation from the continuum level was observed with significance
$>\sigma_{\rm lim}$ standard deviations. Note that the significance of an
absorption feature, $\sigma_n$, is an overestimate because the flux in
adjacent channels is (positively) correlated [each spectral resolution
element is sampled at approximately the Nyquist rate, see Murphy, Curran \&
Webb (2002) for details].

According to the ``expected'' $\delta_{\rm LOS}=100$ \kms ~velocity
differences in the optical and millimetre wave lines-of-sight$^5$, the
only candidate features identified in our analysis are HCO$^+$
$2\rightarrow3$ and CO $3\rightarrow4$ towards B 0458--020 as well as
the HCO$^+$ $2\rightarrow3$ line towards B 0834--201. As seen in Table
4, however, increasing the line-of-sight velocity difference to
$\delta_{\rm LOS}=200$ \kms~ significantly increases the number of
candidates.  This suggests that either the authenticity of the
candidate features is questionable and/or we have
\begin{table*}[hbt]
\caption[]{Candidate molecular absorption systems located within
$\approx\pm200$ \kms ~of the DLA redshift. $\nu_{obs}$ is the observed
frequency, $\sigma_n$ is the significance of the putative
feature, $z_{\rm mm}$ is the absorption redshift and $z_{\rm 21~cm}$
is the redshift of the \HI ~line (see Curran {\rm et~al.} 2002). The
final column gives the offset in \kms ~between the $z_{\rm mm}$ and
the nominal value of $z_{\rm DLA}$, with our estimate of uncertainty
in the DLA redshift$^3$ quoted.}
\label{cand}
\begin{center}
\begin{tabular}{l c  c c r c r}
\hline\noalign{\smallskip}
DLA & Transition & $\nu_{obs}$ [GHz] &$\sigma_n$  & $z_{\rm mm}$ & $z_{\rm 21~cm}$ & $z_{\rm mm}-z_{\rm DLA}$\\
\noalign{\smallskip}
\hline\noalign{\smallskip}
B 0248+430 & CS $2\rightarrow3$ & 105.3701 & 4.2 & 0.39479 & 0.3941 & $190\pm20$\\
B 0458--020 & HCO$^+$  $2\rightarrow3$ & 87.9994 & 3.8 & 2.04045 & 2.03945 &$50\pm10$\\
...& ...& 88.0229 & 3.2 & 2.03964 & ...&$-30\pm10$\\
...& CO $3\rightarrow4$& 151.7144 & 4.3 & 2.03888 & ...& $-100\pm10$\\
B 0738+313 &  CO $0\rightarrow1$ & 94.3116 & 4.4 & 0.22224 & 0.2212 &$240\pm20$\\
B 0834--201 & HCO$^+$  $2\rightarrow3$ & 98.5135 & 4.4 & 1.71595 & --&$100\pm100$\\
\noalign{\smallskip}
\hline\noalign{\smallskip}
PKS 1413+135 & HCO$^+$  $2\rightarrow3$ & 214.6110 & 4.6 & 0.24671 & 0.24671 &--\\
\noalign{\smallskip}
\hline
\end{tabular}
\end{center}
{Note: The last entry is a known absorption system included as a
guide. A spectrum was obtained after $\approx 30$ hours with SEST (August
2001) which may well have been rejected upon visual inspection$^3$.}
\end{table*}
underestimated the uncertainties in the DLA redshifts. It is
interesting to note that all but one of the candidate lines occur in
DLAs with high column densities ($N_{\rm HI}\ga10^{21}$ \scm). The one
exception is B 0834--201 which has the highest visual magnitude of
these candidates. We emphasise that Table 4 is only intended as a
shortlist for follow up observations, which may or may not yield
absorption systems at the listed redshifts.

\begin{figure}[h]
\vspace{6.7 cm} \setlength{\unitlength}{1in} 
\begin{picture}(0,0)
\put(-0.3,3.05){\includegraphics{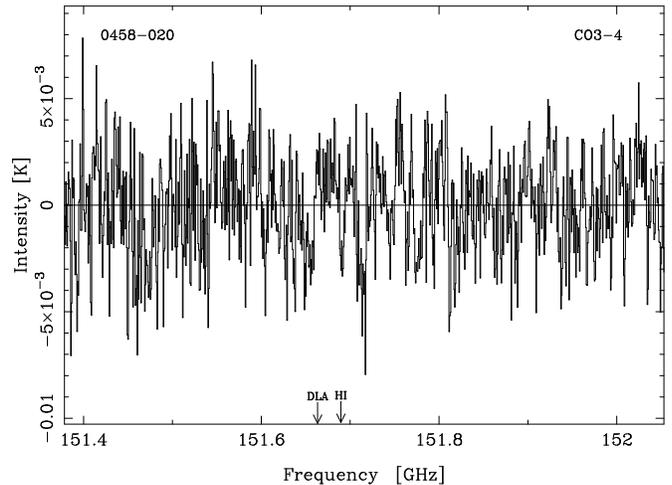}}
\end{picture}
\caption[]{CO $3\rightarrow4$ at $z=2.039$ towards B 0458--020 at the
full spectral resolution of 1.38 \kms. The antenna temperature of the
peak channel is $8$ mK (c.f. $\sigma=3$ mK) and the integrated
main-beam intensity of the line is $-0.10\pm0.02$ K \kms. The expected
frequencies of the line for the (approximate) DLA redshift of 2.0399
and \HI~ redshift of $2.03945$ ({Turnshek} {\rm et~al.} 1989, {Wolfe}
{\rm et~al.} 1995) are shown.}
\label{d1}
\end{figure}

Of all the candidate systems, the most promising and visually striking
example of a possible absorption line is the 151.71 GHz feature
towards B~0458--020 (Fig. \ref{d1}), which has a high visual magnitude
as well as the highest neutral hydrogen column
density of the sample.
Comparing this ``detection'' of -8 mK (0.24 Jy at SEST) with the
estimated flux density at 152 GHz (0.62 Jy, estimated from the
neighbouring values in Table \ref{t1}) gives an optical depth of
$\tau_{\rm mm}\approx0.5$. The $\approx9$ \kms ~FWHM of the line gives
a column density estimate of $N_{\rm CO}\approx8\times10^{16}$ \scm
~or $N_{{\rm H}_{2}}\sim10^{21}$ \scm, i.e. $\sim20$\% of $N_{\rm
HI}$. This compares well with the column densities and their ratios
for the known absorbers, i.e. $10-40\%$ ({Wiklind} \& {Combes} 1994,
1995, 1996c, 1999, {Carilli} {\rm et~al.} 1998), although it is
considerably higher than the rest of the DLA sample.

\section{Summary}

We have observed 18 transitions for molecular absorption over the
redshift range $0.2\ga z \ga3$ in 11 damped Lyman-alpha
absorption systems which lie along the line-of-sight towards
radio-loud quasars. Of these, 9 quasars have flux densities $\ga0.1$
Jy in the millimetre band, making our observations sensitive to
optical depths $\ga0.1$ at a $3\sigma$ level and a spectral resolution
of 10 \kms, thus improving
the limits for 11 transitions over the previous results.

From these observations and the previously published results, we
estimate an upper limit of $\approx2$\% for the molecular hydrogen to
neutral atomic hydrogen column density ratio. This is the lower limit
derived by {Carilli} {\rm et~al.} (1996) based on a molecular hydrogen
detection and a redshifted \HI ~21 cm non-detection. This is
consistent with the results of {Liszt} (2002) who suggests that the
low metallicity of the Lyman-alpha system is not favourable for the
formation of molecules, but our limit perhaps suggests that we are
close to the values necessary to detect absorption lines in at least
the high column density ($N_{\rm HI}\ga4\times10^{21}$ \scm)
DLAs. Subsequently, from an analysis of the spectra we find several
significant features located close to the expected redshifts, which
do, however, require confirmation. Therefore follow up observations of
the candidates given in this article should be the next step in
searching for the elusive high redshift molecular absorption systems.

\begin{acknowledgements}
We wish to thank the John Templeton Foundation for supporting this
work. SJC acknowledges receipt of a UNSW NS Global
Fellowship and FTR  acknowledges support from the Chilean
{\it Centro de Astrof\'\i sica} FONDAP No. 15010003.
\end{acknowledgements}

                                                  

\end{document}